\documentclass[10pt,twocolumn,preprintnumbers,amsmath,showpacs]{revtex4}

\usepackage{graphicx}
\usepackage{dcolumn}
\usepackage{bm}
\usepackage[latin1]{inputenc}
\usepackage{psfrag}
\usepackage{epic}
\usepackage{bm}
\usepackage{latexsym}

\newcommand{\trace}[1]{\mathrm{Tr}\, #1}
\newcommand{\vkt}[1]{{\boldsymbol{#1}}}
\newcommand{\matr}[1]{#1}
\newcommand{\transp}{\mathrm{T}}
\newcommand{\expt}[1]{\langle #1\rangle}
\newcommand{\bra}[1]{\langle #1|}
\newcommand{\ket}[1]{|#1\rangle}
\newcommand{\braket}[2]{\langle #1|#2 \rangle}
\newcommand{\dd}{d}
\newcommand{\expe}{e}
\newcommand{\imi}{\mathrm{i}}

\begin{document}

\title{Entanglement in bosonic systems}

\author{Stein Olav Skr\o vseth}
\email{stein.skrovseth@ntnu.no}
\affiliation{%
Department of Physics,
Norwegian University of Science and Technology,
N-7491 Trondheim, Norway
}%

\date{9 September 2005}

\begin{abstract}
  We present a technique to resolve a Gaussian density matrix and its
  time evolution through known expectation values in position and
  momentum. Further we find the full spectrum of this density matrix
  and apply the technique to a chain of harmonic oscillators to find
  agreement with conformal field theory in this domain. We also
  observe that a nonconformal state has a divergent entanglement
  entropy. 
\end{abstract}

\pacs{03.67.Mn, 11.10.-z,11.25.Hf}

\maketitle

\section{Introduction}
Entanglement is today considered a fundamental resource in nature
when it comes to quantum computation and
information \cite{Nielsen&Chuang}, and measures of entanglement has
become a major field of research. In particular, entanglement in
condensed-matter systems and the entanglement's critical behavior is
well investigated
\cite{Osterloh:2002,Vidal:2002rm,Wei04,SOS1,Verstraete2004}. In 
these terms, the entanglement entropy is an analytically well suited
tool for investigating the properties of ground states in
condensed-matter systems. In this paper, we focus on bososnic states
with Gaussian wave functions, and the entanglement properties of the
ground state of a simple harmonic chain, which belongs to this class
\cite{Plenio2005}.

We consider
the notion of entanglement entropy,
that is, considering a quantum system denoted $\mathcal C$ in a
pure state with
wave function $\ket\psi$, we trace out some portion $\mathcal B$ to
obtain the density matrix of the remaining space $\mathcal A$ as
$\rho_{\mathcal A}=\trace_{\mathcal B}\ket\psi\bra\psi$. Then the
entanglement of $\mathcal A$ with respect to $\mathcal B$ is well
defined by the entropy (measured in ebits) of the reduced density
matrix \cite{Nielsen&Chuang, Bennett:1995tk};
\begin{equation}
  S_{\mathcal A}=-\trace\rho_{\mathcal A}\log_2\rho_{\mathcal A}.
  \label{Sdef}
\end{equation}
The measure of entropy in units of ebits is customary and we will use
logarithms base two throughout the paper.
This procedure is well established, and works well for all cases where
the entire strip $\mathcal C$ is in a pure state, though entanglement
measures for mixed states are still incomplete. Most work has been
focused on this entanglement in spin chains, but we will focus on a
one-dimensional bosonic strip \cite{calabrese-2005-0504,
  Audenaert2002}. 

At critical points in a parameter space we have
scale and translational invariance, and thus expect the theory to be
conformally invariant \cite{Francesco97}, and one can use this fact to efficiently detect
critical systems \cite{SOS1}. As was computed by Holzhey, Larsen, and
Wilczek \cite{Holzhey94}, conformally invariant systems in 1+1 dimension can be
considered as a string of length $\Lambda$ of which we trace out some
fraction $1-\sigma\in[0,1]$, and then the entanglement entropy of the remaining
space with respect to the rest is
\begin{equation}
  S(\sigma)=\frac{c+\bar
    c}6\log\left[\frac\Lambda{\pi\epsilon}\sin(\pi\sigma)\right].
  \label{Holzhey}
\end{equation}
Here $c$ and $\bar c$ are the holomorphic and anti holomorphic charges
respectively. $\epsilon$ is
some cutoff parameter that we will consider arbitrary. When
considering the limit $\sigma\ll1$, the formula reduces simply to
$S\sim\log\sigma\Lambda$, which has been a matter of keen interest
\cite{Vidal:2002rm}. However, 
we will focus on any $\sigma$, in particular, when keeping the
$\Lambda$ constant (\ref{Holzhey}) provides a very specific signature
of a conformally invariant system. Thus this formula presents two
independent (as long as $\epsilon$ is considered arbitrary) signatures
of a finite conformal system. First the logarithmic divergence of the
entropy as $\sigma$ is held constant while $\Lambda$ increases, and
second the characteristic $\log\sin$ signature when $\Lambda$ is
constant and $\sigma$ varies.

Another, more trivial, measure of the entanglement of a reduced
density matrix is the product state identification
\begin{equation}
  E_M=1-\trace\rho_{\mathcal A}^M,\qquad M\geq2,
\end{equation}
which is zero for a product state, and unity for a maximally entangled
state. This measure is equivalent to the Rényi entropy \cite{Jin2004}
and is not well 
suited for much more than to 
single out a pure state, as with increasing $M$ any entangled state
will converge to zero in this measure, 
\[\lim_{M\to\infty}E_M=\left\{\begin{matrix}1\quad&\mbox{entangled state}\\
0\quad&\mbox{product state.}\end{matrix}\right.\]

The density matrix contains all information of any system in a
mixed or pure state, and computing vital physical information on any
such system is mostly determined by the eigenvalues of the density
matrix. Therefore there is a great need to compute these in an
efficient way. In particular, considering a
bosonic system in one spatial dimension in its ground state, it can be
modeled as a harmonic chain. The quantum correlations
as measured by the entanglement are nonzero, as will be shown. Thus
the vacuum itself 
is entangled, which is a highly intriguing result. Our main purpose in
this paper is twofold, first we wish to demonstrate how to compute
this entanglement from the vacuum ground state, and second we wish to
see how and if the conformal invariance identified by Eq. (\ref{Holzhey})
arises in this case. The free boson is known to have a central charge
$c=\bar c=1$, so the prefactor in $S(\sigma)$ reduces to simply $1/3$. 
It turns out that we indeed have a conformally invariant ground
state, as identified by the two conformal signatures provided. As we
will demonstrate in 
Sec. \ref{sec:application}, the entropy diverges in the massless limit, but it
nevertheless seems conclusive that the theory is conformally invariant
below a certain threshold mass.

In this paper we will in Sec. \ref{sec:timeev} compute how a
Gaussian density matrix can be recovered from expectation values in
position and momentum, and how this can be used to compute the time
evolution of the density matrix. This can also be computed in terms
of Weyl representation, though we apply a more explicit representation
here. In Sec. \ref{sec:entmeasures} we
compute the two entanglement measures written in this section from the
Gaussian density matrix, and finally in Sec. \ref{sec:application}
we apply the formulas to the ground state of a harmonic chain.

\section{Gaussian density matrices}
\label{sec:timeev}
\subsection{Recovery of the density matrix from expectation values}
It is known that a Gaussian state in a quantum
harmonic oscillator potential will evolve in such a way that the
Gaussian shape is preserved at all times. However, the different
parameters of the Gaussian state may also evolve in time so that the
exact appearance of the density matrix may be difficult to
predict. Consider a Gaussian density matrix of $N$ particles with
positions $q_i$, 
\begin{align}
  \rho(\vkt{q},\vkt{q}')&=
  \sqrt{\frac{\det\left(\matr{A}'-\matr{C}'\right)}{\pi^D}}\exp\left[-d'_i\left(A'_{ij}-C'_{ij}\right)^{-1}d'_j\right]\nonumber\\
  &\times\exp\Big[-\frac{1}{2}\left(q_iA_{ij}q_j+q'_iA^*_{ij}q'_j\right)+q_iC_{ij}q'_j+d_iq_i\nonumber\\
    &+d^*_iq'_i\Big] 
  \label{DM}
\end{align}
summing over repeated indices. Here $\matr{A}$ is a positive, symmetric $N\times
N$ matrix, while $\matr{C}$ is a Hermitian $N\times N$ matrix, and
$\vkt{d}$ is an $N$-dimensional vector. We use $\matr{A}'$, $\matr{C}'$,
and $\vkt{d}'$ to denote the real parts of $\matr{A}$, $\matr{C}$, and
$\vkt{d}$, respectively. We further denote the imaginary parts such
that $\matr{A}=\matr{A}'+\imi\matr{A}''$, etc. Positions and
momenta are real valued. The matrix $A'-C'$ must be
invertible and positive in order to have an normalizable density
matrix. Also, in order to keep $\rho$ positive, that is that
$\bra{\psi}\rho\ket{\psi}>0$ for any $\ket\psi$, one must have
$A'>0$. Hence follows that even $C$ must be
positive.
Next we define three matrices of variances in position and momentum,
\begin{align}
  \begin{split}
    Q_{ij}&=\expt{q_iq_j}-\expt{q_i}\expt{q_j},\\
    P_{ij}&=\expt{p_ip_j}-\expt{p_i}\expt{p_j},\\
    S_{ij}&=\frac{1}{2}\expt{q_ip_j+p_jq_i}-\expt{q_i}\expt{p_j}.
    \label{QPSdef}
  \end{split}
\end{align}
$\matr{Q}$ and $\matr{P}$ are symmetric, while all are
real. Furthermore, since the system in translationally invariant we
can assume $Q_{ij}=f(|i-j|)$ for some function $f$, and similarly with
the other matrices. Thus $Q$, $P$, and $S$ are Toeplitz matrices,
potentially simplifying numerical computations. Toeplitz matrices are
known to be central also in the study of quantum spin chains \cite{Jin2004}.
These three matrices can be
assumed known in a given model of a 
bosonic quantum system. We refer to the set of variables in the
density matrix as $\Theta=\left\{\matr{A},\matr{C},\vkt{d}\right\}$, and
the expectation value matrices as
$\Xi=\left\{\matr{Q},\matr{P},\matr{S},\expt{\vkt{q}},\expt{\vkt{p}}\right\}$.
A simple count shows that the two sets both have $2D^2+3D$ degrees of
freedom, and thus the two sets may contain equal amounts of information.
The expectation value matrices can be computed for a given density
matrix through a straightforward calculation that computes expectation
values of an operator $\hat{\mathcal O}$ as 
\[\expt{\hat{\mathcal O}}=\trace\hat{\mathcal O}\rho(\vkt q,\vkt
q')=\iint\dd\vkt q\dd\vkt q'\delta(\vkt q-\vkt q')\hat{\mathcal
  O}\rho(\vkt q,\vkt q').\]
This means that the expectation value matrices $\Xi[\Theta]$ can be
  computed in terms of the parameters in the density matrix,
\begin{align}
  \begin{split}
    \matr{Q}&=\frac{1}{2}(\matr{A}'-\matr{C}')^{-1},\\
    \matr{P}&=\matr{A}-(\matr{A}-\matr{C})\matr{Q}(\matr{A}-\matr{C})^\transp,\\
    \matr{S}&=-\matr{Q}(\matr{A}''+\matr{C}''),\\
    \expt{\vkt{q}}&=2\matr{Q}\vkt{d}',\\
    \expt{\vkt{p}}&=-2(\matr{A}''-\matr{C}'')\matr{Q}\vkt{d}'+\vkt{d}''.
    \label{ACtoQPS}
  \end{split}
\end{align}
Note that since $A'-C'$ is invertible, $Q$ is well defined invertible. This
set of matrix equations can then be inverted to yield $\Theta[\Xi]$,  
\begin{align}
  \begin{split}
    \matr{A}'&=\matr{P}+\frac{1}{4}\matr{Q}^{-1}-\matr{S}^\transp\matr{Q}^{-1}\matr{S},\\
    \matr{A}''&=-\frac{1}{2}\left(\matr{S}^\transp\matr{Q}^{-1}+\matr{Q}^{-1}\matr{S}\right),\\
    \matr{C}'&=\matr{P}-\frac{1}{4}\matr{Q}^{-1}-\matr{S}^\transp\matr{Q}^{-1}\matr{S},\\
    \matr{C}''&=\frac{1}{2}\left(\matr{S}^\transp\matr{Q}^{-1}-\matr{Q}^{-1}\matr{S}\right),\\
    \vkt{d}' &=\frac{1}{2}\matr{Q}^{-1}\expt{\vkt{q}},\\
    \vkt{d}'' &=\left(\matr{A}''+\matr{C}''\right)\expt{\vkt{q}}+\expt{\vkt{p}}.
    \label{QPStoAC}
  \end{split}
\end{align}  
We note that $\matr{S}=0$ implies both $\matr{A}$ and $\matr{C}$ to be
real, and that in a state where
$\expt{\vkt{q}}=\expt{\vkt{p}}=0$, $\vkt{d}=0$. This way
it is simple to recover the density matrix given $\Xi$. A
strategy to find the nonlinear time evolution of the density matrix
is sketched in Fig. \ref{fig:strategy}; one would use the fact that the time
evolution of $\Xi$ is much simpler to find than that of
$\Theta$. And since the formulas above enables us to switch between the
two sets in an easy manner, we can take the time evolution via $\Xi$
instead of taking it directly on $\Theta$. 
\setlength{\unitlength}{2mm}
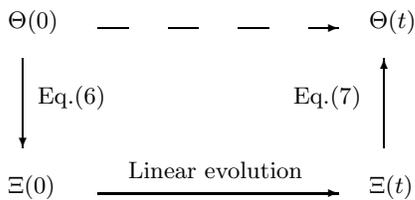
\begin{figure}
  \begin{picture}(30,15)
    \put(1,13){\mbox{$\Theta(0)$}}
    \put(25,13){\mbox{$\Theta(t)$}}
    \put(1,2){\mbox{$\Xi(0)$}}
    \put(25,2){\mbox{$\Xi(t)$}}
    \put(2,11){\vector(0,-1){6}}
    \put(7,2){\vector(1,0){16}}
    \put(26,5){\vector(0,1){6}}
    \put(22,13){\vector(1,0){1}}
    \dashline{2}(7,13)(23,13)
    \put(3,8){\mbox{Eq.(\ref{ACtoQPS})}}
    \put(9,3){\mbox{Linear evolution}}
    \put(20,8){\mbox{Eq.(\ref{QPStoAC})}}
  \end{picture}
  \caption{\label{fig:strategy}Outline of our strategy to find the time evolution of $\Theta(t)$ by going through $\Xi(t)$.}
\end{figure}

\subsection{Time evolution}
The Gaussian shape of the density matrix is preserved under time
evolution governed by the Lagrangian
\begin{equation}
  \mathcal L=\frac12\left(\dot q_n\dot
    q_n-q_n\Omega_{nm}q_m+\xi_nq_n\right),
  \label{KGLagrangian}
\end{equation}
with sum over repeated indices.
The conjugate momentum is $p_n=\dot q_n$, while $\Omega$ is symmetric
and $\xi$ is a real three-vector describing external forces. The equations of
motion become $\ddot q_n+\Omega_{nm}q_m=\xi_n$, which implies four
coupled differential equations 
for the time evolution of the expectation value matrices
\begin{align}
  \begin{split}
    \dot Q&=\mathcal T(S),\\
    \dot P&=-\mathcal T(\Omega S),\\
    \dot S&=P-\frac12Q\Omega,
  \end{split}
\end{align}
where we have defined the symmetrizing operator 
$\mathcal T(A)=A+A^{\mathrm T}$. These
relations specify the time evolution of the system, given some initial
condition. Also, combining the first and third equation gives
\[\ddot Q+2\dot Q-\frac12\{Q,\Omega\}=0,\]
or, in the case of a diagonal $\Omega$,
$\Omega_{ij}=\delta_{ij}\omega_i$, this reads
\[\ddot Q_{ij}+2\dot Q_{ij}-\frac12(\omega_i+\omega_j)Q_{ij}=0.\]

\section{Entanglement measures for Gaussian density matrices}
\label{sec:entmeasures}
The eigenvalues of Eq. (\ref{DM}) can be found explicitly as we will show
in this section. To this end, consider what we will refer to as the
single-particle density matrix
\begin{align}
  \begin{split}
    \rho&^0(x,x';\eta,d)=\expe^{-d'^2/1-\eta}\sqrt{\frac{1-\eta}{\pi}}\,\\
    &\times\exp\left[-\frac{1}{2}\left(x^2+x'^2\right)+\eta xx'+dx+d^*x'\right].
    \label{simpleDM}
  \end{split}
\end{align}
Here $d=d'+\imi d''$ is a complex number, while $\eta$ is real in
the range $[0,1\rangle$. The latter constraint is consistent with the
mentioned requirement on the density matrix that $A'-C$ is positive.
We will omit the two latter parameters in the
argument list of $\rho^0$ when there is no risk of confusion. 
The eigenvalue equation of this density matrix is
\begin{equation*}
  \int\dd x'\rho^0(x,x')\Psi_n(x')=\lambda_n\Psi_n(x).
\end{equation*}
Also, the density matrix obeys
\begin{equation}
  \rho^0(x+x_0,x'+x_0)
  =e^{\imi d''x}\rho^0(x,x';\eta,0)e^{-\imi d'' x'},
  \label{rhod0}
\end{equation}
where $x_0=d'/1-\eta$. This means that the scaled eigenfunction
$\tilde\Psi_n(x)=\expe^{\imi d''(x-x_0)}\Psi_n(x-x_0)$ also is an eigenvector
of $\rho^0(x,x')$ with the same eigenvalue as $\Psi_n(x)$, and hence
the eigenvalues are independent of the displacement $d$. Furthermore,
Eq. (\ref{rhod0}) shows that any traces $\trace\rho^M$ are invariant
under $d$.

Now, to find the eigenvalues of the single-particle density matrix
  when $d=0$, consider the Green's function
  $\mathcal{G}(z,z';\tau)=\bra{z}e^{-\tau 
  H}\ket{z'}$, with $H$ being the single harmonic oscillator
Hamiltonian $H=-p^2/2+\frac{1}{2}\omega z^2$, and the states
$\ket{z}$ the position eigenstates of the quantum harmonic
oscillator. The eigenvalue result for this matrix is, with
$\braket{z}{\Psi}_n=\Psi_n(z)$ and $\ket{\Psi}_n$, the energy
eigenstates
\begin{equation*}
  \int \dd z'\,\mathcal{G}(z,z';\tau)\Psi_n(z')=\expe^{-\omega\tau(n+1/2)}\Psi_n(z).
\end{equation*}
Furthermore, $\mathcal{G}$ must solve the initial value problem
\begin{align*}
  \begin{split}
    \left(-\frac{\partial}{\partial\tau}-H\right)\mathcal{G}(z,z';\tau)=0,\\
    \lim_{\tau\to 0}\mathcal{G}(z,z';\tau)=\delta(z-z'),
  \end{split}
\end{align*}
which, through a scaling of variables,
$x=z\sqrt{\omega\coth(\omega\tau)}$, has the solution
\begin{align}
  \begin{split}
    \mathcal{G}(x,x';\tau)=&\sqrt{\frac{\omega}{2\pi\sinh(\omega\tau)}}\\
    &\times\exp\left[-\frac{1}{2}\left(x^2+x'^2\right)+\frac{xx'}{\cosh(\omega\tau)}\right].
    \label{fullG}
  \end{split}
\end{align}
There also exists a complex solution, which we  disregard as
unphysical. This solution (\ref{fullG}) means that if we identify
$\eta=1/\cosh(\omega\tau)$, we can conclude that $\rho^0$ has the
infinite set of eigenvalues 
\begin{align}
  \label{simpleEV}
  \lambda_n = \sqrt{\omega\coth(\omega\tau)}\,\expe^{-\omega\tau\left(n+1/2\right)}&=\lambda_0\,\xi^n\\
  n&\in\left\{0,1,2,\dots\right\},\nonumber
\end{align}
where $\xi=\expe^{-\omega\tau}=\eta/(1+\sqrt{1-\eta^2})$.

Hence we have identified the eigenvalues and eigenfunctions of our density
matrix in the case $d=0$, which is sufficient to know the eigenvalues
also in the $d\not=0$ case.

Now, let us again turn to our general density matrix (\ref{DM}), and
show that this can be transformed into the simple form of Eq.
(\ref{simpleDM}) \cite{Eisert2003}. In general, the matrix $\matr{A}''$ contributes
nothing to traces of the density matrix, and $A$ can be considered
real. Take an orthogonal matrix $\matr{O}$ 
that diagonalizes $\matr{a}=\matr{O}^\transp\matr{A}'\matr{O}$ and
create the matrix $\sqrt{\matr{a}}$ which is diagonal and takes the
square roots of $\matr{A}$'s eigenvalues on its diagonal. $A$ is
positive by assumption, so this is well defined and real. Then we
find the orthogonal matrix $\tilde{\matr{O}}$ that diagonalizes the
matrix $\bar\eta$ defined as 
\begin{align}
  \bar\eta&=\tilde{\matr{O}}^\transp\sqrt{\matr{a}^{-1}}\matr{O}^\transp\matr{C}\matr{O}\sqrt{\matr{a}^{-1}}\tilde{\matr{O}}.
\end{align}
Note that both $\matr{O}$ and $\tilde{\matr{O}}$ are real, since
$\matr{C}$ is hermitian and $\matr{A}$ is symmetric. 
The coordinates and vectors $\vkt{d}$ transform as
\begin{align*}
  \begin{split}
    \tilde{\vkt{q}}&=\tilde{\matr{O}}\sqrt{\matr{a}}\matr{O}^\transp\vkt{q},\\
    \tilde{\vkt{d}}&=\tilde{\matr{O}}\sqrt{\matr{a}^{-1}}\matr{O}^\transp\vkt{d},
  \end{split}
\end{align*}
which means that the density matrix (\ref{DM}) may be rewritten as
\begin{align}
  \tilde{\rho}(\vkt{q},\vkt{q}')
  &=\prod_{i=1}^N\rho^0(q_i,q_i';\eta_i,\tilde{d}_i),
  \label{rewrittenDM}
\end{align}
$\eta_i$ being the diagonal elements of $\bar\eta$ and
$\tilde{\rho}(\vkt{q},\vkt{q}')=\rho(\tilde{\vkt{q}},\tilde{\vkt{q}}')$.
In other words, the density matrix (\ref{DM}) may be expressed as a
product of single-particle density matrices (\ref{simpleDM}). Thus each of
the product terms in Eq. (\ref{rewrittenDM}) gives the infinite set of
eigenvalues (\ref{simpleEV}), and we may write the eigenvalues for the
total matrix as 
\begin{equation}
  \lambda_{n_1,n_2,\dots,n_N}=\prod_{i=1}^N\lambda_{n_i}=\Lambda_0\prod_{i=1}^N\xi_i^{n_i}.
\end{equation}
Here we have defined $\Lambda_0=\prod_i\lambda^{(i)}_0$.
Here we have let $\lambda_0\mapsto\lambda_0^{(i)}$, $\eta\mapsto 
\eta_i$, and $\xi\mapsto\xi_i$ from Eq.
(\ref{simpleEV}) where the index $i$ refers to particle
number. $\Lambda_0$ can be calculated from the normalization 
condition 
\[\trace{\rho}=\sum_{n_1=1}^\infty\cdots\sum_{n_D=1}^\infty\lambda_{n_1,\dots,n_D}=1\]
giving $\Lambda_0=\prod_{i=1}^N(1-\xi_i)/\xi_i$.
This sets the stage for the calculation of entropy and traces over the
density matrix. 

\subsection*{Entanglement measures}
Having obtained the density matrix eigenvalues, the entropy
$S=-\trace\rho\log\rho$ is easily calculated,
\begin{align}
  S &=-\sum_{n_1=1}^\infty\cdots\sum_{n_N=1}^\infty\lambda_{n_1,\dots,n_N}\ln\lambda_{n_1,\dots,n_N}\nonumber\\
  &=-\sum_{i=1}^N\left[\log\left(1-\xi_i\right)+\frac{\xi_i\log\xi_i}{1-\xi_i}\right],
  \label{S_sum}
\end{align}
a result that agrees exactly with that found by Srednicki \cite{Srednicki93}.

We can proceed to find exact formulas for the entanglement measures
$E_M\equiv 1-\trace\rho^M$ 
where $M\geq2$, for which we find that 
\begin{equation}
  E_M=1-\prod_{i=1}^N\frac{\left(1-\xi_i\right)^M}{1-\xi_i^M}.
  \label{EM_xi}
\end{equation}

These two entanglement measures are now quite straightforward to compute.
Since $Q$ is a Toeplitz matrix, the inversion
involved can be done efficiently with existing linear algebra
packages. Also the computation involves two
diagonalizations of real, symmetric $N\times N$ matrices, which is
also numerically straightforward and efficient. We will focus
primarily on the entropy due to its 
analytical usability.

\section{Application}
\label{sec:application}
We now apply the above formalism to the quantum Klein-Gordon field $\phi_n$,
defined by the Lagrangian (\ref{KGLagrangian}) with $\xi=0$ and
$\Omega_{ij}=\delta_{ij}\kappa$ on $N$
lattice points with periodic boundary conditions
\cite{Mandl&Shaw}. The lattice constant 
is denoted $a$, and the system size is thus $\Lambda=aN$. 
This field has the Fourier expansion in bosonic creation and
annihilation operators $a_{\vkt k}$ and $a^\dag_{\vkt k}$
respectively;
\begin{align}
  \phi_n=\sum_k\frac1{2\Lambda\omega_k}\left(a_k\expe^{-\imi(kn-\omega_kt)}+a_k^\dag \expe^{\imi(kn-\omega_kt)}\right) 
\end{align}
and conjugate field
\begin{equation}
  \pi_n=-\imi\sum_k\sqrt{\frac{\omega_k}{2\Lambda}}\left(a_k\expe^{-\imi(kn-\omega_kt)}-a_k^\dag\expe^{\imi(kn-\omega_kt)}\right).
\end{equation}
In this discrete field theory the dispersion relation is
\[\omega_k^2=\frac4{a^2}\sin^2(k/2)+\kappa^2,\]
 and the sum is over all
allowed wave vectors $k$ in this space. We have the nonzero
commutation rules (at equal time)
\begin{align*}
  [\phi(\vkt{x}),\pi(\vkt{x}')]&=i\delta(\vkt{x}-\vkt{x}'),\\
  [a_\vkt{k},a_\vkt{k}^\dag]&=\delta_{\vkt{k},\vkt{k}'},
\end{align*}
indicating the harmonic oscillator nature of the system, with
positions $\phi_n$ and momenta $\pi_n$.
When rescaling this theory it is imperative to keep $\Lambda$ constant
in order to 
maintain any scaling invariance and hence any conformal invariance in
the model. The $\Xi$-matrices of the ground state are
the Toeplitz matrices
\begin{align}
  \begin{split}
    Q_{mn}&=\expt{\phi_n\phi_m}=\frac{1}{2N}\sum_k\frac{1}{\omega_k}\,\expe^{\imi k(m-n)},\label{Q_n}\\
    P_{mn}&=\frac{1}{2N}\sum_k\omega_k\,\expe^{\imi k(m-n)},\\
    S_{mn}&=0,
  \end{split}
\end{align}
with the sums running over all lattice points numbered $i$, and
$k=2\pi i/N$. $\matr{Q}$ and $\matr{P}$ are thus essentially finite
Fourier transforms of $\omega_k^{\pm1}$. Now, since $\matr{S}$ is
vanishing, we can conclude that both $\matr{A}$ and $\matr C$ are
real. As shown this is all we need to recover the density matrix, and
calculating the traces. 

In the limit of a massless theory, $\kappa\to0$, 
$Q_n$ will diverge due to the $k=0$ (zero mode) term in the
sum (\ref{Q_n}). We can exclude this zero mode from the
sum by summing over lattice point $m$ using $k=\pi(2m+\alpha)/N$
and choose $\alpha=0~(1)$ to compute with (without) the
zero mode. This corresponds to using (anti)periodic boundary
conditions on the field $\phi$.

\begin{figure}
  \psfrag{X}{$\kappa$}
  \psfrag{Y}{$E_2$}
  \psfrag{A1}{ $N=500$}
  \psfrag{A2}{ $N=100$}
  \psfrag{A3}{ $N=25$}  
  \includegraphics[angle=-90]{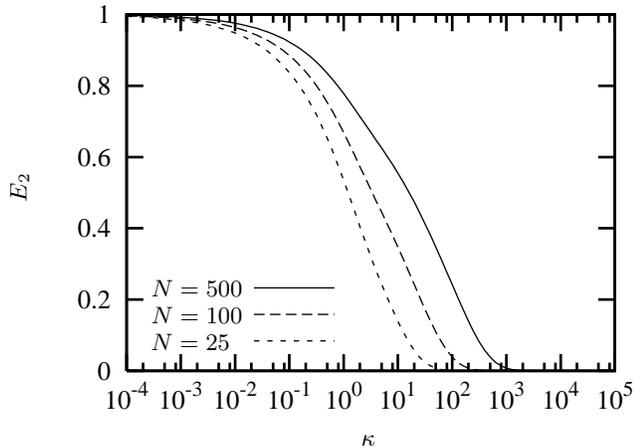}  
  \caption{\label{fig:E2_kappa}$E_2$ as function of $\kappa$ for three
  different system sizes. The transition from an entangled state in
  the massless limit to a product state in the massive limit is
  clearly shown. Here, and in Fig. \ref{fig:S_kappa_a0}, $\kappa$ is in
  units of inverse length $[\phi^{-1}]$.}
\end{figure}

When finding the entropy we define some of the points as ``inside''
(region $\mathcal{A}$) and some as ``outside'' (region $\mathcal{B}$),
and trace over the outside to find the geometric entropy of the inside
region. Formally, this amounts to calculating the $\Xi$ matrices
for the $\sigma N$ inside oscillators, $\sigma$ being the
fraction of the entire system that $\mathcal A$ constitutes.
When the $\Xi$s are known, we can compute the
density matrix and the entropy for the state. Clearly, the entropy is
symmetric with respect to interchange of $\mathcal A$ and $\mathcal
B$, so any entanglement measure will be symmetric around
$\sigma=1/2$. Hence we expect a maximal entanglement at this
half size, and we can consider the half size entanglement with respect
to $\kappa$ and $N$. Results for $E_2$ at half size is shown in Fig.
\ref{fig:E2_kappa}.
Hence we conclude that the massive $\kappa\to\infty$ system in
nonentangled, while we have a transition to a maximally entangled
case in the massless system $\kappa\to0$. Also, the entanglement is
larger for larger systems, and the transition occurs at a larger mass
in a larger system. The correlations are greater in a large system,
and the inertia of the mass $\kappa$ must be larger to prevent them.

\begin{figure}
  \psfrag{X}{$\kappa$}
  \psfrag{Y}{$S_\mathrm{max}$}
  \psfrag{A1}{$N=25$}
  \psfrag{A2}{$N=100$}
  \psfrag{A3}{$N=500$}  
  \includegraphics[angle=-90]{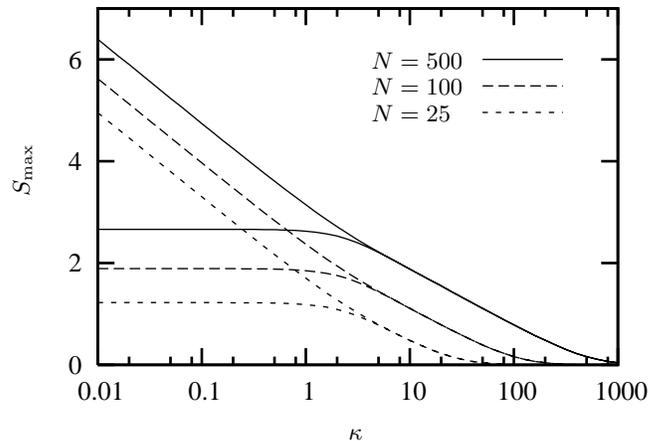}
  \caption{\label{fig:S_kappa_a0}$S_{\mathrm{max}}$ with $\kappa$ for
  the same system sizes as in Fig. \ref{fig:E2_kappa}. The three
  lower lines are $\alpha=1$, while the divergent lines are
  $\alpha=0$. Note that the two lines with different $\alpha$ and same
  $N$ diverge from
  each other at roughly the same $\kappa$ regardless of the system
  size $N$.} 
\end{figure}
\begin{figure}
  \psfrag{X}{$N$}
  \psfrag{Y}{$S_\mathrm{max}$}
  \psfrag{A1}{ $\kappa=10^{-4}$}
  \psfrag{A2}{ $\kappa=10^{-3}$}
  \psfrag{A3}{ $\kappa=10^{-2}$}
  \psfrag{A4}{ $\kappa=10^{-1}$}
  \psfrag{A5}{ $\kappa=1$}
  \includegraphics[angle=-90]{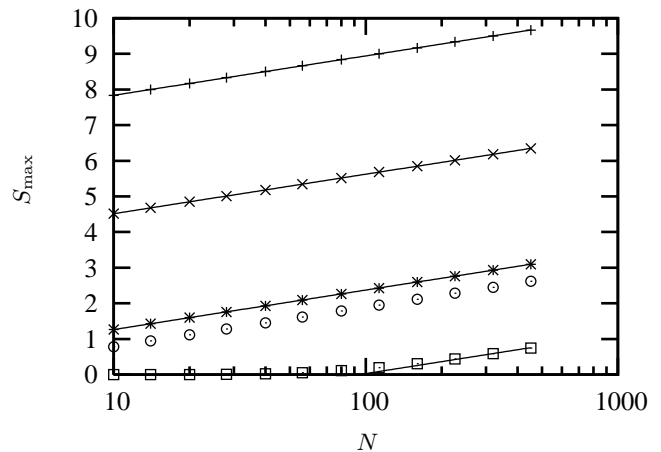}  
  \caption{\label{fig:S_N}$S_\mathrm{max}$ with $N$ for
  various $\kappa$. The straight lines shown all rise with central
  charge $c=1/3$. The bisected points are the $\alpha=0$ points, from
  top to bottom, $\kappa=10^{-4}$, $\kappa=10^{-2}$, $\kappa=1$, and $\kappa=10^2$.
  The circles are for $\alpha=1$ and $\kappa=0$.}
\end{figure}

We now consider the entanglement entropy. The half size measure for
both $\alpha=0$ and $\alpha=1$ are shown in Fig.
\ref{fig:S_kappa_a0}. We see that when the zero mode is included, the
entropy diverges in the massless limit, while without the zero mode it
converges to some system dependent value. In Fig. \ref{fig:S_N} we see
how the half size entropy diverges logarithmically with system size for
small $\kappa$. Most remarkably, this divergence occurs regardless of
the zero mode, and the correct scaling factor $c/3$ is reproduced.

\begin{figure}
  \psfrag{X}{$\sigma$}
  \psfrag{Y}{$S$}
  \includegraphics[angle=-90]{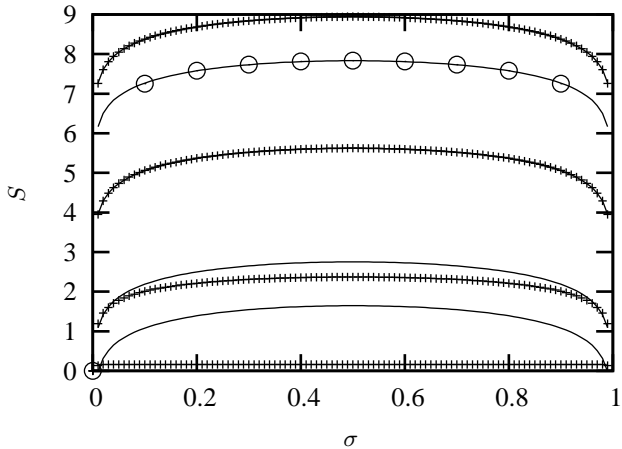}
  \caption{\label{fig:S_s_a0}$S(\sigma)$ for some values of
    $\kappa$ with $\alpha=0$. Here, $N=100$($+$) and $N=10$($\bigcirc$). From top to
  bottom, $\kappa=10^{-4}$, $\kappa=10^{-2}$, $\kappa=1$, and
  $\kappa=10^2$. The full lines are
  $f(\sigma)=\frac{1}{3}\log_2(\sin\pi\sigma)+a$, with
  $a$ chosen to
  fit the lines at the ends. For large $\kappa$, the entanglement
  saturates and does not obey the conformal signature. In the massless
  limit the conformal signature is obeyed, even at $N=10$.} 
\end{figure}

Finally, we look at the second feature of the conformal signature,
namely the $\log\sin$ shape of a finite system. For the
$\alpha=0$ case this is shown in Fig. \ref{fig:S_s_a0}, and we again
see a good characteristic of the conformal system in a massless
system. For the massive system, the entanglement saturates, and fits
the signature only at small $\sigma$, which is also observed earlier
in noncritical quantum spin chains \cite{Vidal:2002rm}.
When the zero mode is omitted, however, the $\log\sin$ signature is not present,
as the system is identical to the $\alpha=0$ case for some massive
$\kappa$, a state that is not conformally invariant. But then it is
nevertheless notable that the logarithmic divergence with system size
still fits the 
conformal theory although the fixed-size signature does not, and the
state is not conformally invariant.

\begin{figure}[t]
  \psfrag{Y}{$\log_{10} s_n$}
  \includegraphics[angle=-90]{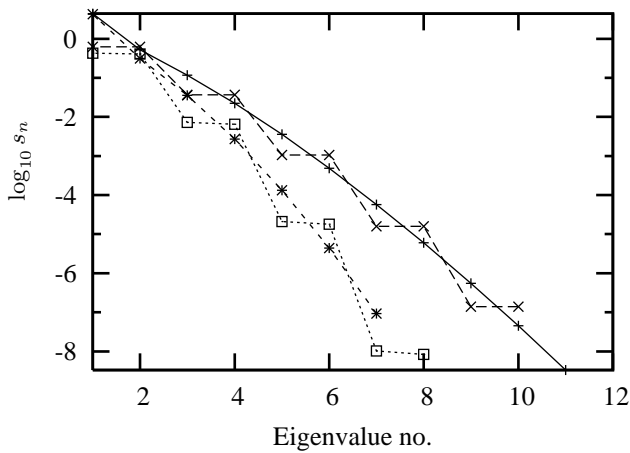}
  \caption{\label{fig:eigvals}The largest entropy contributions in the
  sum $S=\sum_ns_n$, with ordered terms. The graph shows the terms for
  system size $N=100$, $\alpha=0$ ($+$), and $\alpha=1$ ($\times$)
  and also system size $N=25$, $\alpha=0$ ($*$), and $\alpha=1$
  ($\Box$). All values are for an essentially massless theory with $\kappa=10^{-3}$. }
\end{figure}
It is valuable to note that the conformal signature in the
entanglement entropy is present even in small systems, such as
$N=10$. This has been seen in quantum spin chains earlier,
there enabling efficient identification of criticality \cite{SOS1}.

As a final feature, we investigate which modes, or terms, contribute to
the entanglement entropy in the expansion $S=\sum_ns_n$ where the
terms are defined by the sum (\ref{S_sum}). The individual terms are
shown in Fig. \ref{fig:eigvals}, where we see that for the
conformally invariant case, the eigenvalues fall of faster than
exponentially, which indicates that only very few terms in the sum are
needed to compute the entropy. 
Indeed, to compute the entropy to
within an error $\pm10^{-6}$ one needs only eight out of a possible 50
terms in the expansion. Also, the figure shows that for the
nonconformal state without the zero mode, the eigenvalues are paired
since in a not scale invariant, though translationally invariant
state, any mode with a nonzero impulse will be degenerate with
another state of opposite impulse.

\section{Conclusion}
We have shown how to compute the entanglement entropy of a chain of
bosonic harmonic oscillators, using scalings and rotations of the
density matrix to put it in a single-particle form. The results show
what we call conformal signatures in the massless limit. Moreover, the
results show that even a nonconformal state can show a logarithmic
divergence as predicted by conformal field theory, but not the
$\log\sin$ signature that the author believes to be a uniquely
conformal feature.

\section{Acknowledgments}
The author wants to thank Professor K\aa re Olaussen for initiating the
research and providing invaluable guidance. The Department of Physics
at the University of Troms\o,
is thanked for their hospitality.

\bibliography{phys}

\end{document}